\documentclass[twocolumn,amsmath,amssymb,superscriptaddress,pra,aps]{revtex4-1}

\usepackage{epsfig}
\usepackage{color}
\usepackage{graphicx}
\usepackage{bm}
\usepackage{epstopdf}
\usepackage[T1]{fontenc}

\begin{document}

\title{Impenetrable Mass-Imbalanced Particles in One-Dimensional Harmonic Traps}

\author{A.~S. \surname{Dehkharghani}}
\affiliation{Department of Physics and Astronomy, Aarhus University,
DK-8000 Aarhus C, Denmark} 
\author{A.~G. \surname{Volosniev}}
\affiliation{Department of Physics and Astronomy, Aarhus University, DK-8000 Aarhus C, Denmark}
\affiliation{Institut f{\"u}r Kernphysik, Technische Universit{\"a}t Darmstadt, 64289 Darmstadt, Germany}
\author{N.~T. \surname{Zinner}}
\affiliation{Department of Physics and Astronomy, Aarhus University,
DK-8000 Aarhus C, Denmark} 

\date{\today}

\begin{abstract}
Strongly interacting particles in one dimension subject to external confinement have become a 
topic of considerable interest due to recent experimental advances and the development 
of new theoretical methods to attack such systems. In the case of equal mass fermions or 
bosons with two or more internal degrees of freedom, one can map the problem onto the well-known 
Heisenberg spin models. However, many interesting physical systems contain mixtures of particles
with different masses. Therefore, a generalization of the recent strong-coupling techniques would 
be highly desirable. This is particularly important since such problems are generally considered
non-integrable and thus the hugely successful Bethe ansatz approach cannot be applied.
Here we discuss some initial steps towards this goal by
investigating small ensembles of one-dimensional harmonically trapped particles where pairwise interactions are either vanishing or infinitely strong with focus on the mass-imbalanced case. 
We discuss a (semi)-analytical approach to describe systems using hyperspherical coordinates where the interaction is effectively decoupled from the trapping potential. 
As an illustrative example we analyze mass-imbalanced four-particle two-species mixtures with strong interactions between the two species. For such systems we calculate the energies, densities and pair-correlation functions. 
\end{abstract}

\maketitle

\noindent
\section{Introduction.}
One-dimensional (1D) quantum systems have recently 
been realized experimentally in highly controllable 
environments using cold atomic 
gases \cite{paredes2004,kino2004,kinoshita2006,haller2009,serwane2011,gerhard2012,wenz2013,
murmann2015a,murmann2015b}. 
The one-dimensional geometry is typically achieved 
with specially designed optical lattices \cite{bloch2008}
and the Feshbach resonance technique \cite{chin2010} is used 
to modify the scattering length between particles, which in
turn changes effective one-dimensional interactions \cite{olshanii1998}. 
An important point in this respect is 
that due to the low temperature and diluteness of the 
system these interactions can be assumed to be of zero range to a very 
high level of accuracy.

Experimental access to systems with tunable interactions 
allows us to witness realizations of theoretical 
models that were considered unrealistic in the past. 
For instance, the Tonks-Girardeau \cite{tonks1936,girardeau1960,olshanii1998} 
and super-Tonks-Girardeau \cite{astrak2005,batchelor2005,tempfli2008} Bose gases 
have been engineered recently \cite{paredes2004,kino2004,haller2009}. Another groundbreaking 
achievement has been the ability to prepare few-body 
fermionic systems in one-dimensional geometries \cite{serwane2011}, 
which drives the theoretical investigation of mesoscopic 
ensembles. In this regard, it was shown that a one-dimensional 
harmonically trapped two-component fermionic few-body systems 
undergo a transition from a non-magnetic to a magnetic phase 
by increasing the strength of repulsion \cite{gharashi2013,bugnion2013,jon2014,cui2014}. 
For the ground state of a Fermi sea with an impurity, 
this transition pushes the strongly interacting impurity 
into the middle of the trap \cite{jon2014, jesper2014}. 
At the same time the opposite behavior is seen for an impurity 
in an ideal Bose gas where the impurity is mostly found at 
the edges of the trap \cite{dehkharghani1, dehkharghani2}. 
This implies interesting quantum magnetic properties of the 
Bose mixtures \cite{dehkharghani1,deuret2014}.
Both of these phenomena can be understood using hyperspherical 
coordinates where the interaction is decoupled from the 
trap and the properties of the ground state are determined 
just by geometrical means \cite{dehkharghani1}.

The main objective of the present paper is to explore these 
geometrical ideas for particles with different masses. Therefore, our study
continues the investigation of few-body mass-imbalanced 
systems, which are known to show properties significantly 
different compare to equal mass cases. For instance, the Efimov 
scenario in three dimensions can be affected in various ways 
using different atomic species \cite{hammer2006}, mass difference 
also changes three-body universally in two and in one dimension, 
see, e.g., Refs.~\cite{filipe2013, dodd1972,kart2008}.

Before we proceed with our presentation we relate the present work
to some previous studies in the literature. For mixtures of fermions
and bosons with equal mass, the technique of Bose-Fermi mapping has
been around for a while \cite{girardeau2004,girardeau2007}, 
but these mappings will in general not generate eigenstates 
that can be adiabatically extended to large but finite interaction 
strengths \cite{volosniev2014a}. An exceptional case is the 
three-body problem \cite{deuret2008a} where the Bose-Fermi mapping will
work as long as the external trap has reflection symmetry. 
In the case of Bose-Fermi mixtures for equal mass 
particles, the Bose-Fermi mapping can be misleading in some
cases \cite{fang2011}. This can be fixed by applying the 
results of Ref.~\cite{volosniev2014a}. Also, it is possible to 
present mathematical arguments on how to use symmetries of 
the problem to determine the eigenstates for both 
non-interacting and strongly interacting equal mass particles
\cite{harshman2012,harshman2014,garciamarch2014,harshman2015a,harshman2015b}. 
These arguments are, however, not extendible to mass-imbalanced
systems at this point. 
The case of mass-imbalanced systems has been addressed for small 
trapped systems in a handful of very recent works
\cite{mehta2014,dehkharghani2,pecak2015,mehta2015}. For instance,
a phase separation in the light component of Fermi-Fermi mixtures has
been reported \cite{pecak2015} and similar results have been found for 
a heavy impurity in an ideal Bose gas of light mass particles \cite{dehkharghani2}.
For particles in a hard-wall box, an exceptional exact solution was recently found 
in the case of four hard-core particles with masses $m$, $2m$, $3m$ and $6m$ \cite{olshanii2015}. 
This was accomplished by mapping onto a geometric problem related to an octacube.

\begin{figure}[t]
\centering
\includegraphics[width=\columnwidth]{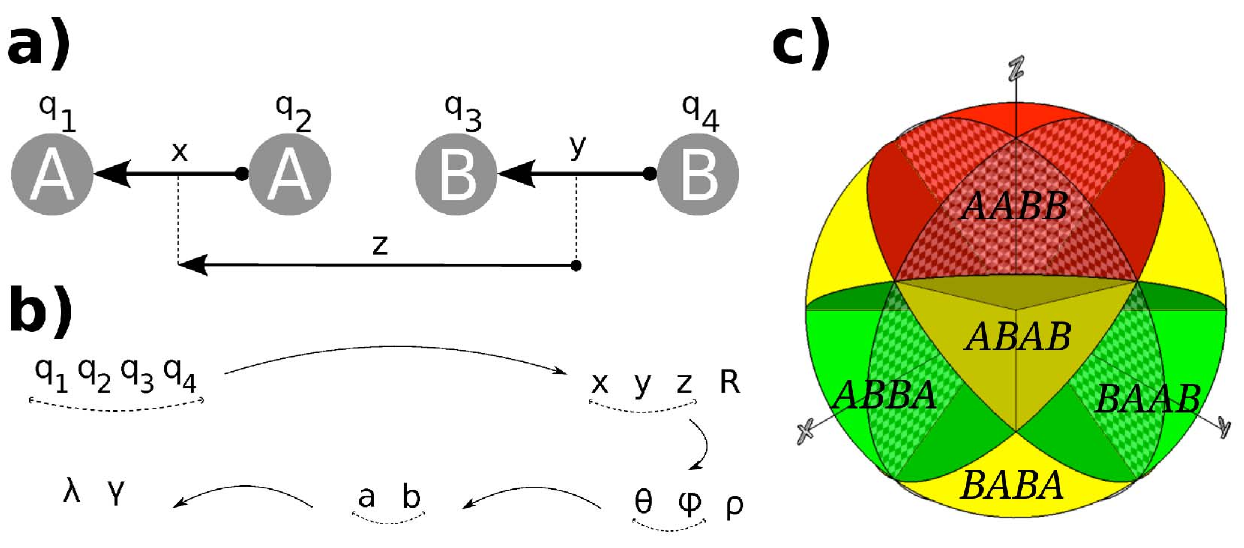}
\caption{{\bf a)} Overview of the $q_i$-coordinates used in the four-body system and the corresponding relative Jacobi coordinates. The intrinsic relative coordinates of the $AA$ and the $BB$ pairs are called $x$ and $y$, respectively, while the $z$-coordinate is used to describe the relative motion of the centers of mass of the two pairs. Note, that the total center-of-mass coordinate, $R$, is not shown. {\bf b)} The transformations used in the paper to solve four-body problems, see the text for details. {\bf c)} The coordinate space for the relative motion with the $\delta$-function interaction planes. The regions are correspondingly divided into different configurations as labeled in the figure.
The transparent checkered planes, $x=0$ and $y=0$ are the planes where the intra-species (among the same species) interactions take place, which are assumed to be small and hence ignored.}
\label{fig1}
\end{figure}

In this paper we develop an approach, { where the geometry of the coordinate space of all configurations is very crucial}. This allows us to study the effect of a mass imbalance on the properties of a trapped strongly interacting one-dimensional system. First we present a general formulation for the $N$ particle one-dimensional system in a harmonic trap assuming that every pair of particles is either non-interacting or infinitely repulsive. { As a warm-up, we present some considerations regarding the two- and three-body cases.} Afterward we provide derivations for the two-species four-body systems { the so-called 2+2 system}. In the case of species with equal mass (see Fig.\ref{fig1}) this system has been investigated in Ref.~\cite{dehkharghani1} (see also Ref.~\cite{garcia2013} and Ref.~cite{garcia2014}). Thus, we focus on the case where particles have the same trapping frequencies but different masses. We provide numerically exact solutions for the two-component systems where each species has two particles that are either fermions or bosons. { In particular, we find a transition between two different types of ordering for the two bosons and two fermions system when in fact the fermions are approximately 1.3 times heavier than the bosons}. In addition, we provide two appendices with technical details, and also an appendix that discusses the related case of three identical fermions and a single impurity of a different mass.

\section{Formulation}
Let us consider a harmonically trapped system with $N$ particles that have masses $\{m_i\}_{i=1}^N$ and coordinates $\{q_i\}_{i=1}^N$. We assume that the interaction between the $i$th and $j$th particles is of zero range, i.e., $V_{ij}=g_{ij}\delta(q_i-q_j)$. In this paper we study the limit where the system has either infinite or zero coupling coefficients, $g_{ij}$. For clarity, at the beginning we fix $1/g_{ij}=0$ for all $i$ and $j$. At the end of this section we discuss what will be different if instead some particles are non-interacting. We start by emphasizing that since the masses are different the Bose-Fermi mapping \cite{girardeau1960,girardeau2004,girardeau2007} cannot be applied. However, the problem can still be treated in a simpler way compared to the system with finite values of $g_{ij}$. This seemingly counter intuitive observation also holds true for free particles on a line, see, e.g., Ref.~\cite{mcguire1972}. 

Let us start by writing the Hamiltonian that describes the system
\begin{equation}
\mathcal{H}=\sum_{i=1}^N\left(-\frac{\hbar^2}{2m_i}\frac{\partial^2}{\partial q_i^2}+\frac{m_i\omega^2 q_i^2}{2}\right),
\end{equation}
where $\omega$ is the trap frequency. The interaction terms are cast into the { boundary} condition that the wave function $\Psi$, which solves $\mathcal{H} \Psi = E \Psi$ for $q_i\neq q_j$, vanishes if $q_i=q_j$. Throughout this paper we assume that this wave function is normalized, i.e., $\int\left|\Psi\right|^2 \mathrm{d}q_1...\mathrm{d}q_N =1$. To proceed further, we notice that the center-of-mass is decoupled, see, e.g., Ref. \cite{jeremy2011,jeremy2012}, from the boundary conditions and the relative coordinates, i.e., $\mathcal{H}=\mathcal{H}_{CM}+\mathcal{H}_{r}$ and $\Psi=\phi_{CM}\psi$. This decoupling can be easily noticed in the following set of variables $\tilde q_i= q_i-Q_N$ for $i=1,2,...,N-1$ supplemented by $\tilde q_N=Q_N$ where the center-of-mass is defined as $Q_N=\sum_i {m_i q_i}/M$, assuming that $M=\sum_i m_i$ is the total mass of the system. 

To proceed further we transform the Hamiltonian using a set of Jacobi coordinates $Q_i$
($Q_i=\sum_j U_{ij}q_j$, where $U$ is an orthogonal matrix) in which the motion of the center-of-mass is governed by the following Hamiltonian 
\begin{equation}
\mathcal{H}_{CM}=-\frac{\hbar^2}{2M}\frac{\partial^2}{\partial Q_N^2}+\frac{M\omega^2 Q_N^2}{2},
\end{equation}
whereas the relative motion is described by
\begin{equation}
\mathcal{H}_{r} = \sum_{i=1}^{N-1}\left(-\frac{\hbar^2}{2\mu}\frac{\partial}{\partial Q_i^2}+\frac{\mu\omega^2 Q_i^2}{2}\right),
\end{equation}
where $\mu$ is some arbitrary mass scale. One can prove that such a transformation exists by constructing it. This can be done as follows: First we find the corresponding transformation for two particles -- $Q_1=\frac{\mu_{12}}{\sqrt{\mu}}\left(q_1-q_2\right)$ and $Q_2=\left({\frac{m_1}{M_{12}}}q_1+\frac{m_2}{M_{12}}q_2\right)$, where $M_{12}=m_1+m_2$ and $\mu_{12}=\sqrt{m_1m_2/M_{12}}$. { Note that the coordinates $\{Q_i\}$, similarly to $\{q_i\}$, have units of length. After this point, the proof goes by induction. Indeed, let us assume that the transformation exists for the $n$-particle system. Then for the $(n+1)$-body system we first apply the transformation for a cluster with $n$ particles, next we apply the established two-body transformation for the center-of-mass of the cluster and the last particle.} Notice that the center-of-mass coordinate can be decoupled in various ways and depending on situation other matrices $U$ might be used. We exemplify it while considering the four-body case.

The wave functions for the center-of-mass part are the well-known one-dimensional normalized harmonic oscillator states
\begin{equation}
\phi^{(\chi)}_{CM} = \frac{1}{\sqrt{2^\chi\,\chi!}} \left(\frac{M\omega}{\pi\hbar}\right)^{\frac{1}{4}} e^{
- \frac{M\omega Q_N^2}{2\hbar}} H_\chi\left(\sqrt{\frac{M\omega}{\hbar}}Q_N \right),
\end{equation}
where $ \chi = 0,1,2,\dots$, and $H_\chi$ is the $\chi$th Hermite polynomial. The corresponding energies are $E^{(\chi)}_{CM}=\hbar\omega(\frac{1}{2}+\chi)$.
Since the center-of-mass part is easily solved and essentially trivial we ignore it from now on and work only with the Hamiltonian for the relative motion. 

{ Recall that if $1/g_{il}=0$ the wave function should vanish if $q_i-q_l=0$.} { In the new set of variables $q_i=\sum_{j}U^{-1}_{ij} Q_j$, where $U^{-1}$ is the matrix inverse to $U$. Therefore, the boundaries where the wave function vanishes are given by} 
\begin{equation}
\sum_{j}(U^{-1}_{ij}-U^{-1}_{lj})Q_j=0.
\label{eq:bound_cond_Q}
\end{equation}
{ Let us use the spherical $N-1$ dimensional coordinate system, where the radial coordinate is $\rho=\sqrt{\sum_{i=1}^{N-1} Q_i^2}$, and the rest of the coordinates $\phi_1,...,\phi_{N-2}$ are defined from the following equations } 
\begin{align*}
Q_1=\rho \cos(\phi_1), \nonumber\\
Q_2=\rho \sin(\phi_1)\cos(\phi_2), \nonumber \\
...,\nonumber \\
Q_{N-1}=\rho\sin(\phi_1)...\sin(\phi_{N-2}),
\end{align*}
where $\phi_{N-2}\in[0,2\pi)$ and the other angles range over $[0,\pi]$. { Using these definitions in Eq. ({\ref{eq:bound_cond_Q}}) we see that the boundary conditions are given by the angular coordinates alone, because the radial coordinate factors out.}
The $\rho$-independent boundary conditions suggest to transform the Hamiltonian to spherical coordinates,
\begin{equation}
\mathcal{H}_{r}=-\frac{\hbar^2}{2\mu}\left(\frac{1}{\rho^{N-2}}\frac{\partial}{\partial \rho}\rho^{N-2}\frac{\partial}{\partial \rho}+\frac{\Delta_{S^{N-2}}}{\rho^2}\right)+\frac{\mu\omega^2\rho^2}{2},
\end{equation}
where $\Delta_{S^K}$ is the spherical Laplacian, known also as the Laplace-Beltrami operator on the $K$-sphere. It is natural to look for the eigenfunctions of the Hamiltonian in the following form 
\begin{equation}
\psi=R(\rho)f(\phi_1,...,\phi_{N-2})\sqrt{\rho^{2-N}}, 
\end{equation}
where 
we assume that $f$ solves the following equation 
\begin{equation}
\Delta_{S^{N-2}} f_{\alpha} = \left(\alpha-\frac{(2-N)(N-4)}{4}\right)f_{\alpha},
\label{eq:angular}
\end{equation}
{ where $\alpha$ is a real number},
supplemented by the $\rho$-independent boundary conditions. At the same time the radial wave function, $R(\rho)$, should satisfy the following equation 
\begin{align}
-\frac{\hbar^2}{2\mu}\left(\frac{\partial^2}{\partial \rho^2}+\frac{\alpha}{\rho^2}\right) R_{\alpha} +\frac{\mu\omega^2\rho^2}{2}R_{\alpha} = E_{\alpha} R_{\alpha},
\end{align}
where $E_{\alpha}$ is the energy of the relative motion.
This equation is well known, see, e.g., Ref. \cite{calogero1969}, and allows physically acceptable solutions written as
\begin{equation}
R_{n,\alpha}(\rho)= A_{n,\alpha}\left(\frac{\rho}{\sigma}\right)^{\gamma+1/2}e^{-\frac{\rho^2}{2\sigma^2}}L_n^\gamma\left(\frac{\rho^2}{\sigma^2}\right)\; ,
\end{equation} 
where $L_n^a$ denotes the generalized Laguerre polynomials, $\sigma\equiv\sqrt{\hbar/(\mu\omega)}$, $2 \gamma\equiv\sqrt{1-4\alpha}$ and $A_{n,\alpha}$ is a normalization constant. The corresponding energy has the following form
\begin{equation}
E_{n,\alpha}=\hbar\omega(2n+\gamma+1)\; ,
\end{equation}
where $n=0,1,2,...$ is the radial quantum number.
It is interesting to note that after we performed these transformations we clearly see that the problem of $N$ harmonically trapped strongly interacting particles in 1D can be mapped onto a problem of finding the 'electrostatic' potential in $N-1$ spatial dimensions for a perfectly conducting wedge filled with an angle-independent charge density. 

In general it is hard to find an analytical solution to Eq.~\eqref{eq:angular}. Even though sometimes it can be solved (e.g., if $m_i=m_j$ for all $i$ and $j$, it can be addressed using the Bose-Fermi mapping \cite{girardeau1960}) for general mass ratios one needs numerical investigation of the problem. As an illustrative example of such a study, we investigate four-body systems below, where the equation can be solved using standard computational techniques. Once the solutions to Eq.~\eqref{eq:angular} are found one can study the properties of the system at infinite interaction strength. Furthermore, the energy to order $1/g_{ij}$ can be also found by applying the Hellmann-Feynman theorem, see, e.g., Refs. \cite{volosniev2014, dorte2015} for equal-mass studies ({ The energy in this order allows one to calculate Tan's contact.}). At the same time, if the angular function $f$ is degenerate, which happens for example if some of the masses are the same, then the Hellmann-Feynman theorem can be used to lift the degeneracy. This allows one to find the state at $1/g=0$ that is adiabatically connected to the ground state of the system for large but finite $g_{ij}$, see, e.g., Ref. \cite{volosniev2014a}.
Moreover, even without finding an explicit solution one can extract important information about the properties of the system. For instance, one can deduce the ordering of the particles in the ground state by comparing solid angles that different orderings possess. This will be illustrated below for four-body systems. 

As we have seen the trapping potential completely decouples from the interaction. It is worthwhile noting that this fact can be related to a hidden $SO(2,1)$ symmetry in the system \cite{pitaevski1997, castin2004, castin2006, sergej2012}. A straightforward consequence of this is that by changing the frequency in time, i.e., $\omega=\omega(t)$ for $t>0$, we only affect the quantum numbers of the radial part. Moreover, since the time-dependent one-body harmonic oscillator is solvable \cite{husimi1953}, we can write down a solution to the corresponding time dependent equation
\begin{equation}
\psi(t)=N(t) \sqrt{\rho^{2-N}} e^{i \mu \rho^2 \frac{\dot\lambda}{2\hbar \lambda}}R\left(\frac{\rho}{\lambda(t)}\right)f(\phi_1,...,\phi_{N-2}),
\end{equation} 
where $R(\rho)f(\phi_1,...,\phi_{N-2})\sqrt{\rho^{2-N}}$ is the initial eigenstate that corresponds to the energy $E$, and
\begin{align}
\frac{1}{N}\frac{\mathrm{d}N(t)}{\mathrm{d}t}=- \frac{1}{2\lambda}\frac{\mathrm{d}\lambda(t)}{\mathrm{d}t}-i\frac{E}{\hbar \lambda^2}, \\
\lambda^3 \frac{\mathrm{d}^2 \lambda(t)}{\mathrm{d}t^2}=\omega^2(t=0)-\omega^2\lambda^4,
\end{align}
supplemented with the initial conditions $\lambda(t=0)=1$ and $\mathrm{d}\lambda/\mathrm{d}t(t=0) =0$. Moreover, $N(t)=1$ for an initially normalized state.
The presented formulas provide us with a unique possibility to study the transition probabilities for a many-body system as was done for a one-body system in Ref.~\cite{popov1969}. However, this investigation is out of scope of the present paper and is left for future studies.

It is important to note that even though at the beginning we stated that $1/g_{ij}=0$ for all $i$ and $j$, the approach remains valid also if some of the $g_{ij}$ vanish. The only difference is that in this case the planes where the 
$i$th and $j$th particles meet are no longer special for the angular wave function $f$. { In other words, the angular equation to solve is the same, however the number of boundary conditions is changed.}

\section{Two- and Three-body Systems}
After the general discussion above we now exemplify by first considering the simplest case of two strongly interacting particles. The Hamiltonian for this system is 
\begin{equation}
\mathcal{H}=\sum_{i=1}^2\left(-\frac{\hbar^2}{2m_i}\frac{\partial^2}{\partial q_i^2}+\frac{m_i \omega^2 q_i^2}{2}\right)+g_{12}\delta(q_1-q_2).
\end{equation}
Applying the two-body transformation, discussed in the previous section, we obtain 
\begin{equation}
\mathcal{H}=-\frac{\hbar^2}{2\mu}\frac{\partial^2}{\partial Q_1^2}+\frac{\mu\omega^2 Q_1^2}{2}+\frac{g_{12}\mu_{12}}{\sqrt{\mu}}\delta(Q_1) + \mathcal{H}_{CM} \; .
\end{equation}
Notice that such a Hamiltonian allows us to find the eigenspectrum for any value of $g_{12}$ Indeed, 
similar to the equal mass case \cite{busch1998,idzia2006}, we derive the following even parity wave function for the relative motion 
\begin{equation}
\psi(Q_1)=A e^{-\frac{Q_1^2}{2\sigma^2}}U\left(-\nu,\frac{1}{2},\frac{Q_1^2}{\sigma^2}\right) \; ,
\end{equation}
where $\nu=\frac{E}{2\hbar\omega}-\frac{1}{4}$ is found from the following equation
\begin{equation}
\frac{\Gamma(-\nu+1/2)}{\Gamma(-\nu)}=-\frac{g_{12}\mu_{12}}{2 \sigma\hbar\omega\sqrt{\mu}}\; ,
\end{equation}
where $\Gamma(x)$ is the gamma function. Note that the odd parity states have wave functions that vanish whenever two particles are at the same position. Hence, in this case the system is effectively non-interacting and the eigenstates are given by the odd parity non-interacting wave functions. It is worth noting that the simple treatment we present here rests on the two assumptions that are crucial for this paper, i.e., the zero-range interaction and the same trapping frequency for two atoms. If these assumptions are not met the correlations in the system are more complicated, see for instance Refs.~\cite{blume2008, deuret2008} for the relevant studies of three-dimensional set-ups.

Unlike the two-body case, three-body systems with finite interactions do not allow simple analytical solutions in general. In the case of three dimensions a number of papers have explored numerical solutions of the harmonically trapped 
three-body problem \cite{luu2007,kastner2007,liu2010,gharashi2012}.
Indeed, in the three-body case by transforming the Hamiltonian to spherical coordinates one will see that the boundary conditions couple $\rho$ and $\phi$ coordinates if $0<g_{ij}<\infty$ \cite{harshman2012, jon2014, volosniev2014}. At the same time if the coupling coefficients, $g_{ij}$, are either vanishing or infinitely large, then the problem again becomes very simple, since in this case Eq.~\eqref{eq:angular}
is one-dimensional. This in turn allows one to obtain a number of results for three-body systems in a rather simple 
manner \cite{nikolaj2014, nielsjacob2015}.

\section{Four-body systems}
Having discussed the two- and three-body cases that allow analytical treatment if $1/g_{ij}=0$ we turn to our key example -- a four-body system. The steps taken to solve this problem are in principle the same as for any system with $N>4$. However, the case $N=4$ can be visualized geometrically, see Fig.~\ref{fig1}. For clarity we consider two-species systems with strong inter-species and vanishing intra-species interactions. 
Bosonic species are denoted $A$ and $B$, while fermionic species are denoted $\uparrow$ and $\downarrow$ to conform to the usual convention of identifying internal components of fermionic system with a spin arrow.
Notice that the theory can be easily extended to multi-component systems (more internal states or more species like $A$, $B$ and $C$ for bosons and so on). { One can solve the problem once the right coordinates are taken in full generality and then pick the solutions that fit the symmetry one is interested in, which will then depend on the quantum statistics of the particles involved.} To describe the positions of particles we introduce the coordinates $\{q_i\}_{i=1}^4$, where the first $N_{A(\uparrow)}$ coordinates describe $A/\uparrow$ particles, i.e., for the 2+2 systems (in the $N_{A/\uparrow}+N_{B/\downarrow}$ notation) $q_1, q_2$ are coordinates of the two $A/\uparrow$ particles and $q_3, q_4$ are for $B/\downarrow$ particles, 
see Fig.~\ref{fig1}{\bf a)}. We also adopt harmonic oscillator units, i.e., we measure lengths in units of $\sigma=\sqrt{\hbar/(\mu\omega)}$, where for concreteness we take $\mu=({m_1m_2m_3m_4/(m_1+m_2+m_3+m_4)})^{(1/3)}$, and energies in units of $\hbar\omega$. Accordingly, the Hamiltonian can be written as
\begin{align}
\mathcal{H}&=\frac{1}{2} \sum_{i=1}^{4}\left(\frac{\mu}{m_i}p_i^2+\frac{m_i}{\mu} q_i^2\right) \;.
\end{align}
This Hamiltonian needs to be supplemented with the interaction term, $\sum_{i=1,i<j}^{4}g_{ij}\delta(q_i-q_j)$, which defines the boundary conditions of the system at the points where two interacting particles overlap.

Following the discussion in the previous section, we first decouple the center-of-mass part using the set of coordinates $(x,y,z,R)$ defined as follows (see, e.g., Ref.~\cite{mcguire}) $\mathbf{J}\mathbf{q}=\mathbf{r}=(x,y,z,R)^T$ with $\mathbf{J}\in O(4)$ and
\begin{align*}
\mathbf{J}=
\frac{1}{\sqrt{\mu}} \begin{bmatrix}
\mu_{12} & -\mu_{12} & 0 & 0 \\
0 & 0 & \mu_{34} & -\mu_{34} \\
\mu_{12,34}\frac{m_1}{M_{12}} & \mu_{12,34}\frac{m_2}{M_{12}} & -\mu_{12,34}\frac{m_3}{M_{34}} & -\mu_{12,34}\frac{m_4}{M_{34}} \\
\frac{m_1}{\sqrt{M}} & \frac{m_2}{\sqrt{M}} & \frac{m_3}{\sqrt{M}} & \frac{m_4}{\sqrt{M}}
\end{bmatrix},
\end{align*}
where $M_{ij}=m_i+m_j$ is the mass of two atoms, the total mass, as before, is denoted with $M$, and 
$\mu_{ij}=\sqrt{{m_im_j}/{M_{ij}}}$ and $\mu_{ij,kl}=\sqrt{{M_{ij}M_{kl}}/{M}}$. 
The volume element will change in the following way 
\begin{equation}
\mathrm{d}q_1\mathrm{d}q_2\mathrm{d}q_3\mathrm{d}q_4\rightarrow\sqrt{\frac{\mu}{M}}\mathrm{d}x \mathrm{d}y \mathrm{d}z \mathrm{d}R. 
\end{equation}
The transformation allows us to write the Hamiltonian in the simple form
\begin{align}
\mathcal{H}=&\frac{1}{2}(-\Delta+\mathbf{r}^2),
\end{align}
where the Laplacian is denoted by $\Delta$. The planes where the particles meet, i.e., $q_i=q_j$, are specified by
\begin{align}
&q_1=q_2\to x=0, \qquad q_3=q_4 \to y=0, 
\label{eq:bound_cond_xyz1}\\
&q_1=q_3 \to z=\frac{m_2\mu_{12,34}}{M_{12}\mu_{12}}x-\frac{ m_4 \mu_{12,34}}{M_{34}\mu_{34}}y,\\
&q_1=q_4 \to z=\frac{m_2\mu_{12,34}}{M_{12}\mu_{12}}x+\frac{ m_3 \mu_{12,34}}{M_{34}\mu_{34}}y,\\
&q_2=q_3\to z=-\frac{m_1\mu_{12,34}}{M_{12}\mu_{12}}x-\frac{ m_4 \mu_{12,34}}{M_{34}\mu_{34}}y,\\
&q_2=q_4 \to z=-\frac{m_1\mu_{12,34}}{M_{12}\mu_{12}}x+\frac{ m_3 \mu_{12,34}}{M_{34}\mu_{34}}y.
\label{eq:bound_cond_xyz6}
\end{align}
After decoupling the center-of-mass motion we end up with the intrinsic motion part described by the $x,y,z$ coordinates. Notice that the wave function of the relative motion, $\psi$, is supplemented by the normalization condition, 
\begin{equation}
\int\left|\psi\right|^2 \sqrt{\frac{\mu}{M}}\mathrm{d}x \mathrm{d}y \mathrm{d}z=1. 
\end{equation}
To proceed we need to specify on which boundaries the wave
function should vanish, i.e., we should specify for each particle
in our system whether it is a boson or a fermion.

\subsection{2+2 system} 
\begin{figure*}[t]
\centering
\includegraphics[width=\textwidth]{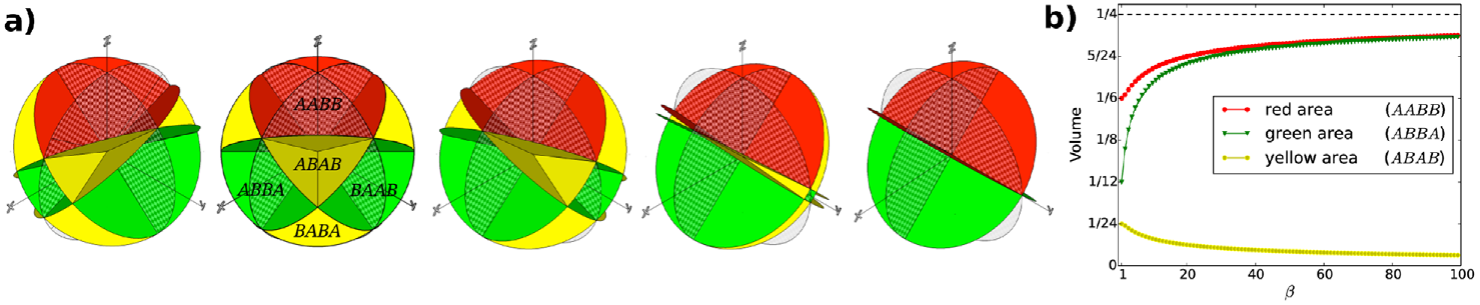}
\caption{\textbf{a)} The geometry of Jacobi-space for the relative motion in the 2+2 system for different values of the mass ratio -- $\beta=0.2, 1, 5, 100$ and $\infty$ from left to right, where $\beta=m_3/m_1$ (in practice we keep $m_1$ fixed and vary $m_3$). For the convenience of the reader we also depict orderings of particles using a bosonic labeling $A$ and $B$ -- for fermions we would use $\uparrow$ and $\downarrow$ instead. Notice that the (green) $BAAB$ area shrinks while the (green) $ABBA$ area grows, as $\beta$ grows. At the same time, notice how the (yellow) areas corresponding to $ABAB$ or $BABA$ orderings are disappearing. \textbf{b)} Volume fractions (volume of the space for a given ordering divided by the total volume for some reference sphere) as a function of $\beta$. The volumes of the $AABB$, $ABBA$, and $ABAB$ configurations are plotted. Notice that for $\beta=1$ the volume-fractions are $1/6$, $1/12$ and $1/24$ for the red, green and yellow areas respectively. The $AABB$ (green) and $ABBA$ (red) volume fractions both converge to $1/4$, while the $ABAB$ (yellow) volume fraction vanishes as $\beta\rightarrow\infty$.}
\label{space}
\end{figure*}

We now consider the main example, which is the 2+2 system, 
i.e., two particle of one species and two particles of another species. {However, another combination such as the 3+1 system is also interesting to investigate and this is briefly discussed in Appendix~\ref{app:A}.}
Here the intra-species interactions are set to zero and the 
inter-species interactions are assumed to be equally strong. 
This implies that $g_{12}=g_{34}=0$ and $g_{13}=g_{14}=g_{23}=g_{24}\to\infty$. 
We visualize the geometry for the relative 
motion in { Fig.~\ref{fig1} and Fig.~\ref{space}, which show the ordering of the particles}. The wave function should 
vanish on the solid planes in Fig.~\ref{space}, whereas identical 
particles overlap on the checkerboard patterned planes. { Notice that these planes are nothing more than just the conditions set by Eq. (\ref{eq:bound_cond_xyz1})-(\ref{eq:bound_cond_xyz6}). Deducing the ordering of the particles can be done by analyzing the different combinations that one can have by ordering the particles on a line, for instance: $q_1<q_2<q_3<q_4$ and conclude which divided space the ordering is possible. Of course this is easily seen on Fig. \ref{fig1} or Fig. \ref{space} as one can visualize}. 
It is important to notice that only the mass ratio, $\beta=m_3/m_1$, 
enters in the boundary conditions making it the only relevant parameter of the problem.
Already by a naked-eye inspection of the volumes of different wedges in Fig.~\ref{space}, we can learn about the structure of the ground state. For example, for $\beta=1$ the volume of the (red) $AABB$ area is much larger, but shares a similar shape to the other regions. In fact, the (red) $AABB$ region can be built by combining two volumes of the (green) $ABBA$ regions, or four of the (yellow) $ABAB$ regions. We can therefore conclude that the ground state of a two-species bosonic system should be located in the (red) 
$AABB$ region. Furthermore, the parity symmetry demands the wave function to have the $AABB\pm BBAA$ configuration, see also Ref.~\cite{dehkharghani1}. 
Note that, as the Hamiltonian conserves parity every state is always a superposition of parity partners. For convenience, in what follows we will always consider only one of these partners unless explicitly stating otherwise.

To proceed further we define spherical coordinates, $\rho$, $\phi$ and $\theta$ as: $x=\rho~\sin\theta \cos\phi$, $y=\rho~\sin\theta \sin\phi$, $z=\rho~\cos\theta$, where $\rho\in[0,\infty[$, $\phi\in[0,2\pi[$ and $\theta\in[0,\pi]$. 
In these coordinates, the Hamiltonian for relative motion becomes
\begin{align}
\mathcal{H}_r=\frac{1}{2}\left(\rho^2-\frac{1}{\rho^2}\frac{\partial}{\partial \rho}\!\left(\rho^2 \frac{\partial}{\partial \rho}\right)-\frac{\Delta_{S^2}}{\rho^2}\right),
\end{align}
where 
\begin{align}
\Delta_{S^2}=\frac{1}{\sin\theta}\frac{\partial}{\partial \theta}\!\left(\sin\theta \frac{\partial}{\partial \theta}\right)
\!+\!\frac{1}{\!\sin^2\theta}\frac{\partial^2}{\partial \phi^2} \; .
\end{align}
{ Using the spherical coordinates in Eqs. (\ref{eq:bound_cond_xyz1}) - (\ref{eq:bound_cond_xyz6}) we obtain the following $\rho$-independent boundary conditions} 
\begin{equation}
\psi(\cos\theta\pm \sin(\phi \pm \xi)\sin\theta)=0,
\end{equation} 
where $\xi\equiv \mathrm{atan}(\sqrt{\beta})$. Note that if the masses are the same, i.e., $\beta=1$, we have $\xi=\pi/4$. 

Following the prescription of the previous section, we separate the wave function into the radial, $R$, and the angular, $f$, parts. The radial part is simple and to describe the system we need to find the angular part, which solves the equation
\begin{equation}
\Delta_{S^2}f_{\alpha}=\alpha f_{\alpha}, 
\end{equation}
supplemented by the condition 
\begin{equation}
f_{\alpha}(\cos\theta\pm \sin(\phi \pm \xi)\sin\theta)=0. 
\end{equation}
This problem is quite straightforward to approach. 
For instance, to solve it in the (red) $AABB$ area we apply two subsequent transformations 
\begin{align}
&i)\,a=\cos\phi~\tan\theta,\,b=\sin\phi~\tan\theta\quad\textrm{and}&\\
&ii)\,\lambda=a~\sin\xi - b~\cos\xi,\, \gamma=a~\sin\xi + b~\cos\xi. &
\end{align}
These transformations cast the boundary conditions in a very simple form, 
so a decomposition of $f_{\alpha}$ in the sine basis can be used to solve the problem, 
see Appendix \ref{app:B} for more details. For clarity we sum up the 
transformations that were performed in Fig.~\ref{fig1}{\bf b)}.

\begin{figure}[t]
\centering
\includegraphics[width=\columnwidth]{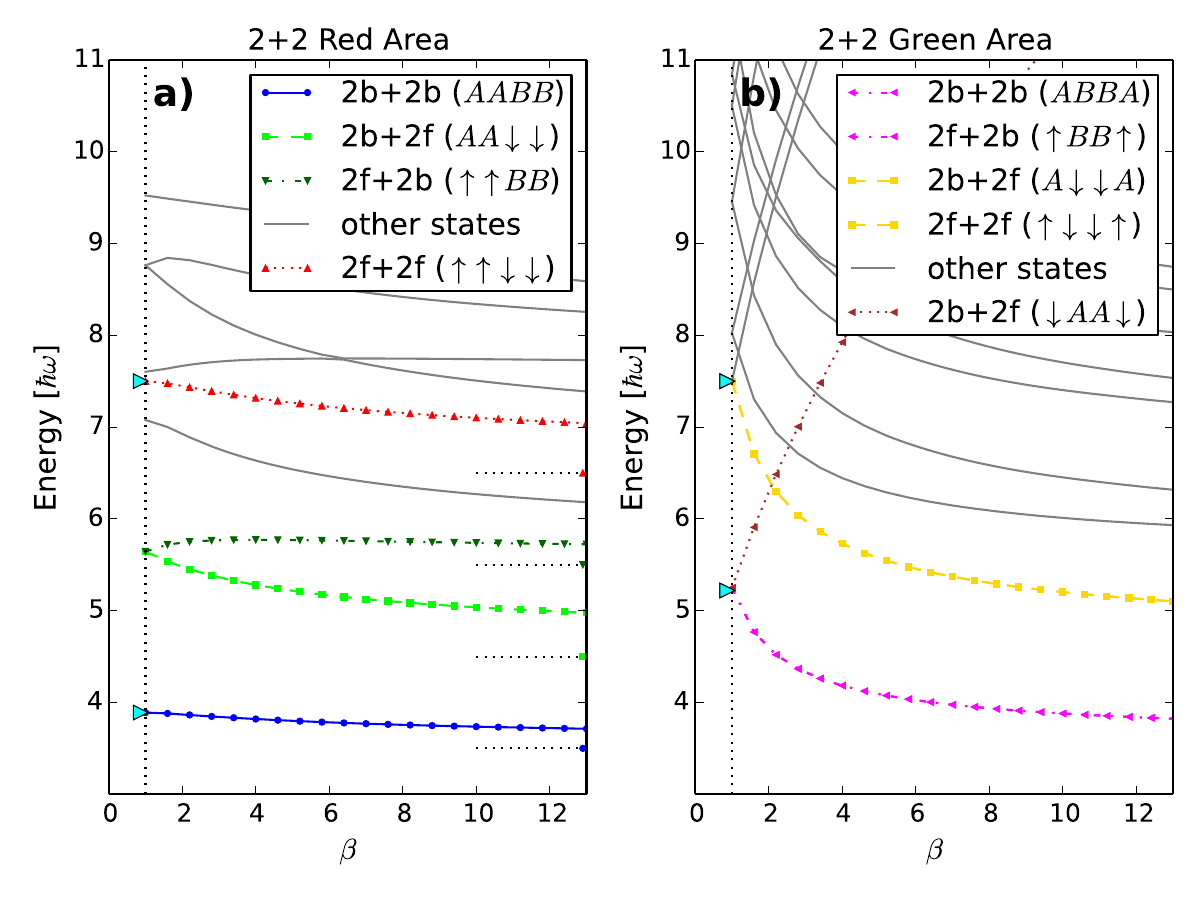}
\caption{The energies for the 2+2 
systems as a function of the mass ratio, $\beta$. The total
center-of-mass energy has been subtracted. {\bf a)} Solutions
of the eigenvalue problem for wave functions that are non-zero
in the (red) area with ordering $AABB$ (for Bose-Bose mixtures). 
The legend shows different orderings with letters denoting
bosonic and arrows denoting fermionic species.
Panel {\bf b)} is similar to panel {\bf a)} but now for the wave functions that
live in the (green) area with ordering $ABBA$. On the same panel we present the $\downarrow AA\downarrow$ configuration, see the text for details.
Note that for the latter case the energies grow very fast 
due to the diminishing volume as $\beta$ increases. 
For some states we also show the asymptotic values for $\beta\rightarrow\infty$ with 
dotted lines. The
triangles at $\beta=1$ are known values from previous numerical studies 
\cite{jon2014, dehkharghani1}. The (colored) lines 
with symbols correspond 
to the four lowest states of a given symmetry, one of which 
should be the global ground state. The thin (grey) lines
correspond to higher excited states.}
\label{energies_as_mass}
\end{figure}

After we solve the angular equation in every domain we have the energies and the corresponding wave functions. In this subsection we would like to address properties of the ground states for particles of different symmetries. Hence we introduce a shorthand notation where, for instance, 2b+2f denotes a system where the first two particles are bosons and the second pair is fermionic, etc. This gives us four possible combinations: 2b+2b, 2b+2f, 2f+2b and 2f+2f. 
Note that in this subsection for each of these systems we will consider only the $\beta\geq 1$ situation. Recall from above that $\beta$ was defined as 
$\beta=m_3/m_1$, thus the mass of the particle mentioned last divided by that mentioned first. For example, in the 2b+2f case, $\beta$ is 
the fermion mass divided by the boson mass. The case with $\beta<1$ follows immediately from the four cases by symmetry 
(2b+2f and 2f+2b are simply interchanged).

Assuming that the radial quantum number is zero, $n=0$, we plot the energy of relative motion in Fig.~\ref{energies_as_mass}. These energies are obtained by solving the angular equation in the largest (red) $AABB$ and (green) $ABBA$ areas. At the same time, by looking at the symmetries of the obtained angular wave functions we determine which systems (bosonic, fermionic) are described with these functions. Since we are after the ground state properties we disregard the (yellow) $ABAB$ area, which is in any case small. For the same reason we present our calculations for the other green region only for the 2b+2f system, see the discussion below. The lowest energy states for each of these systems are labeled in Fig.~\ref{energies_as_mass} for both the (red) $AABB$ and (green) $ABBA$ areas together with the $\downarrow AA \downarrow$ configuration. By comparing the energies from the panels we can find the global ground state for a given system. The identification of the ground state is easy for the 2b+2b, 2f+2f and 2f+2b, because these states are favored in a specific ordering for all $\beta$. For the 2b+2f case we observe an intriguing change of configuration as a function of mass, which we will discuss below. 

\begin{figure}[t]
\centering
\includegraphics[width=\columnwidth]{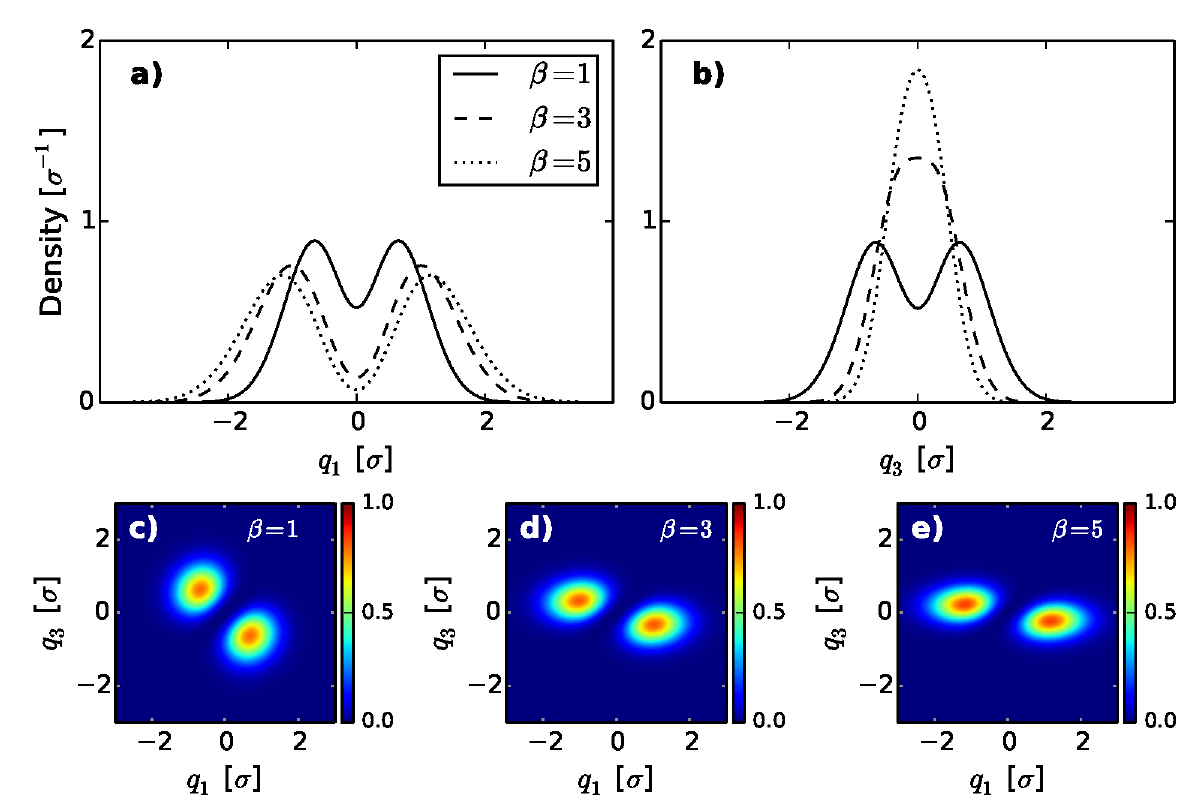}
\caption{(color online). The densities and pair-correlation functions for the ground state ($AABB\pm BBAA$ configuration) of the 2b+2b system. Panels {\bf a)} and {\bf b)} show the densities for the $A$ and $B$ particles, respectively. The mass ratios are $\beta=1,3$ and $5$, i.e., the $B$ particles are one, three and five times heavier than the $A$ particles. Panels {\bf c)}, {\bf d)} and {\bf e)} show the corresponding $AB$ pair-correlation functions.}
\label{density_2b+2b}
\end{figure}

\subsubsection{2b+2b} 
We first consider a bosonic system. Here we have large interspecies and vanishing intraspecies interactions which is a regime that 
has previously been denoted the composite fermionization regime \cite{zollner2008}. It was shown in Ref.~\cite{dehkharghani1} that if all masses are the same the configuration $AABB$ (or equivalently $BBAA$) minimizes the energy. For the mass-imbalanced case as follows from Fig.~\ref{energies_as_mass} the ground state also corresponds to the $AABB$. In fact, the ground state wave function consists of an equal weight superposition of the $AABB$ and $BBAA$ configurations due to the parity invariance of the external trap. This can be easily understood 
from Fig.~\ref{fig1}{\bf c)}, where the (red) $AABB$ area has the largest volume. As the mass ratio increases the lowest energy of the $ABBA$ configuration rapidly decreases, see Fig.\ref{energies_as_mass}{\bf b)}. However, the energy of the $AABB$ ordering also decreases, yet slower, but enough to stay the ground state configuration. Notice that these two configurations become degenerate if one of the particles is infinitely heavy. This can be understood by considering a harmonic oscillator with an impenetrable wall (of zero width) in the middle which represents the two infinitely heavy $B$ particles sitting on top of each other at the origin. The ground state is four-fold degenerate since the two mobile particles can be placed in four different ways; the two $A$ particles can both sit on either side of the wall in the lowest eigenstate of the harmonic oscillator that vanishes at the origin (which is the first excited single-particle state of the oscillator), or they may sit on the opposite sides of the wall in a symmetric or antisymmetric combinations.
It is easy to calculate that $7/2\hbar\omega$ is the energy of relative motion in this limit, and the figure shows that this limit is slowly approached. The densities and corresponding pair-correlation functions (see Appendix~\ref{app:C}) for the ground state are shown in Fig.~\ref{density_2b+2b}. One can easily see how the light particles split as the heavy particles move to the center of the trap.

\begin{figure}[t]
\centering
\includegraphics[width=\columnwidth]{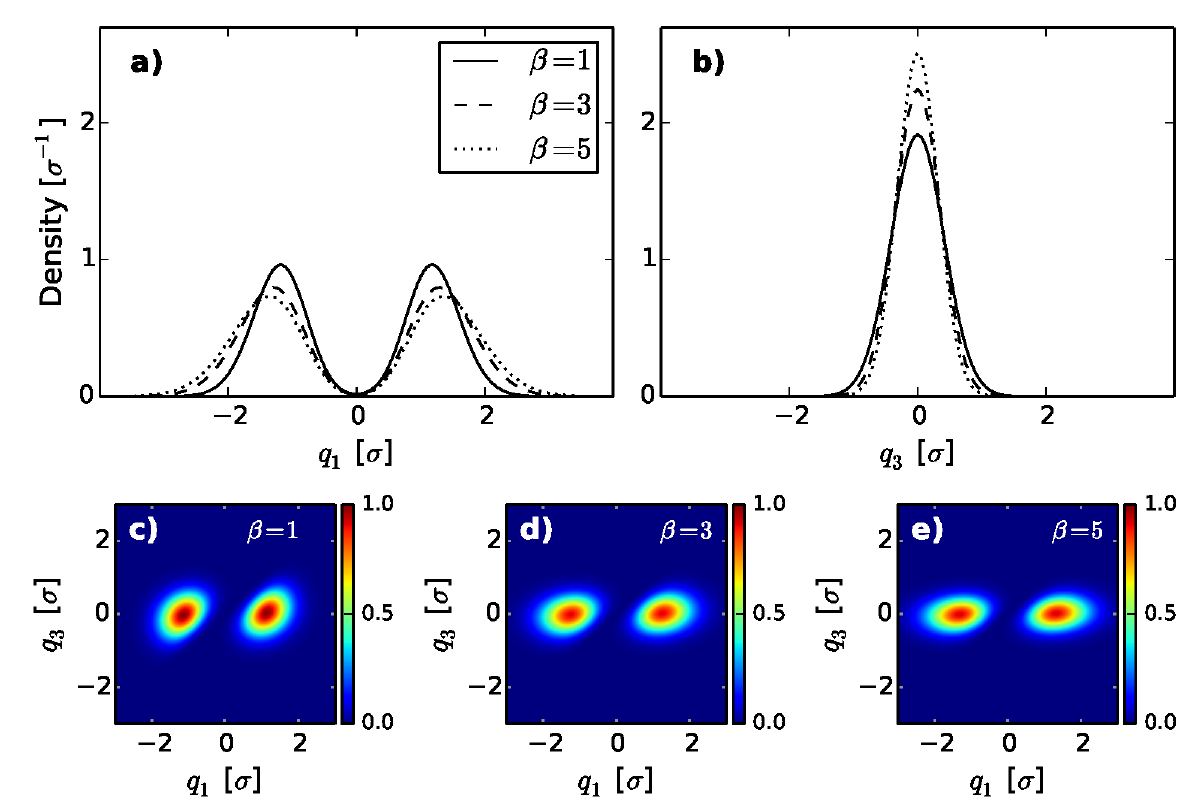}
\caption{(color online). The densities and pair-correlation functions for the ground state ($\uparrow BB\uparrow$ configuration) of the 2f+2b system. Panels {\bf a)} and {\bf b) }show the densities for the $\uparrow$ and $B$ particles, respectively. Again $\beta=1,3$ and 5 is used. Panels {\bf c)}, {\bf d)} and {\bf e)} show the corresponding $\uparrow B$ pair-correlation functions.}
\label{density_2f+2b}
\end{figure}

\subsubsection{2f+2b} 
Next we consider systems with two fermions and two bosons. The system has been analyzed both in the $\uparrow\uparrow BB$ and 
$\uparrow BB\uparrow$ configurations, see Fig.~\ref{energies_as_mass}{\bf a)} and \ref{energies_as_mass}{\bf b)}, respectively. Upon an inspection of these figures it becomes clear that the ground state should have the $\uparrow BB\uparrow$ configuration. The density for the ground state is shown in 
Fig.~\ref{density_2f+2b}. We see that the lightweight fermions are clearly split and the heavier bosons are concentrated in the middle of the trap.

\begin{figure}[t]
\centering
\includegraphics[width=\columnwidth]{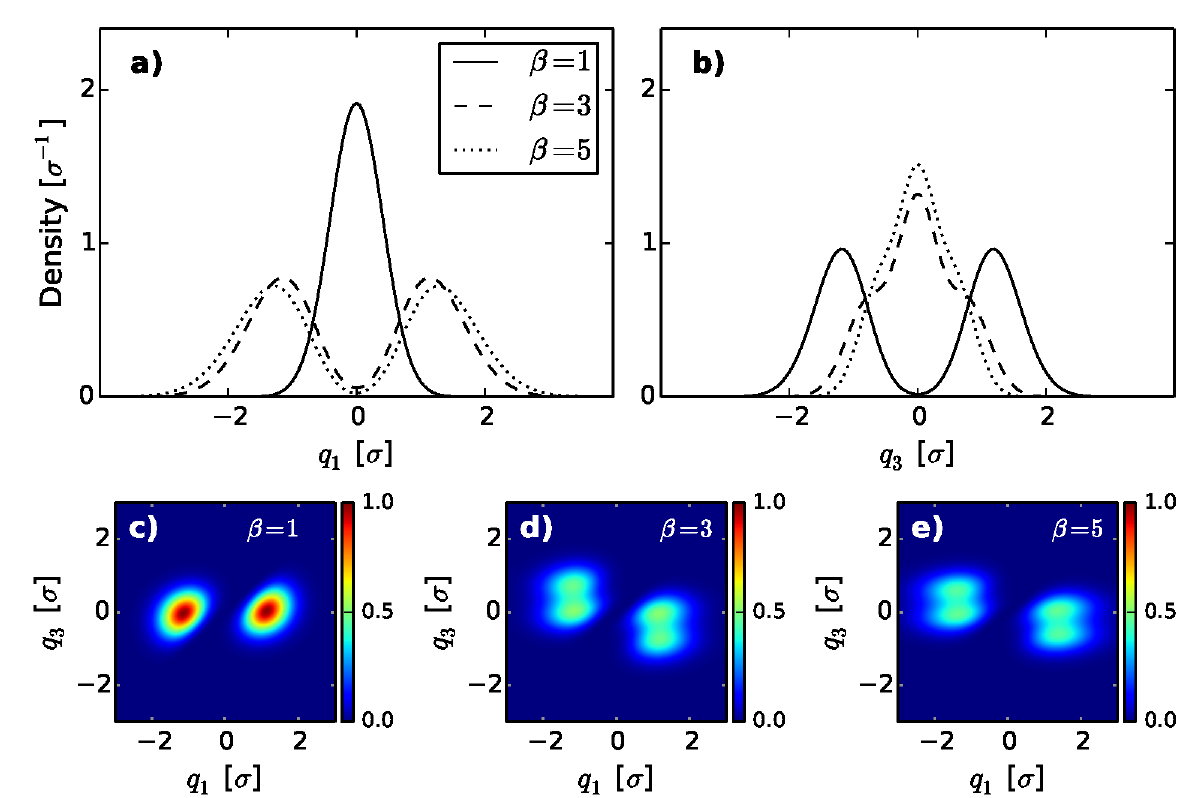}
\caption{(color online). The densities and pair-correlation functions for the ground state ($\downarrow AA\downarrow$ or $AA\downarrow\downarrow \pm \downarrow\downarrow AA$ configurations) of the 2b+2f system; $\beta_c\simeq 1.3$ is found to be the critical value where the ordering in the ground state goes from $\downarrow AA\downarrow$ to $AA\downarrow\downarrow \pm \downarrow\downarrow AA$. Panels {\bf a)} and {\bf b)} show the density for the $A$ and $\downarrow$ particles, respectively. We used $\beta=1,3$ and $5$. Panels {\bf c)}, {\bf d)} and {\bf e)} show the corresponding $A\downarrow$ pair-correlation functions.} 
\label{density_2b+2f}
\end{figure}

\subsubsection{2b+2f} 
The 2b+2f system on the other hand has a configuration depending on the mass ratio. A closer look at the energy spectrum reveals that for $1<\beta <\beta_c $, where $\beta_c\simeq1.3$ (within numerical error), the $\downarrow AA\downarrow$ combination is the lowest state. On the other hand, for $\beta>\beta_c$ the $AA\downarrow\downarrow$ configuration is favored. This transition is the result of the interplay between two important factors. First is the Pauli principle, which pushes the fermions from one another, second is the external trap, which due to the mass difference, pushes the fermions to the center stronger than the bosons. The corresponding density plots for this case are shown in Fig.~\ref{density_2b+2f}.

\begin{figure}[t]
\centering
\includegraphics[width=\columnwidth]{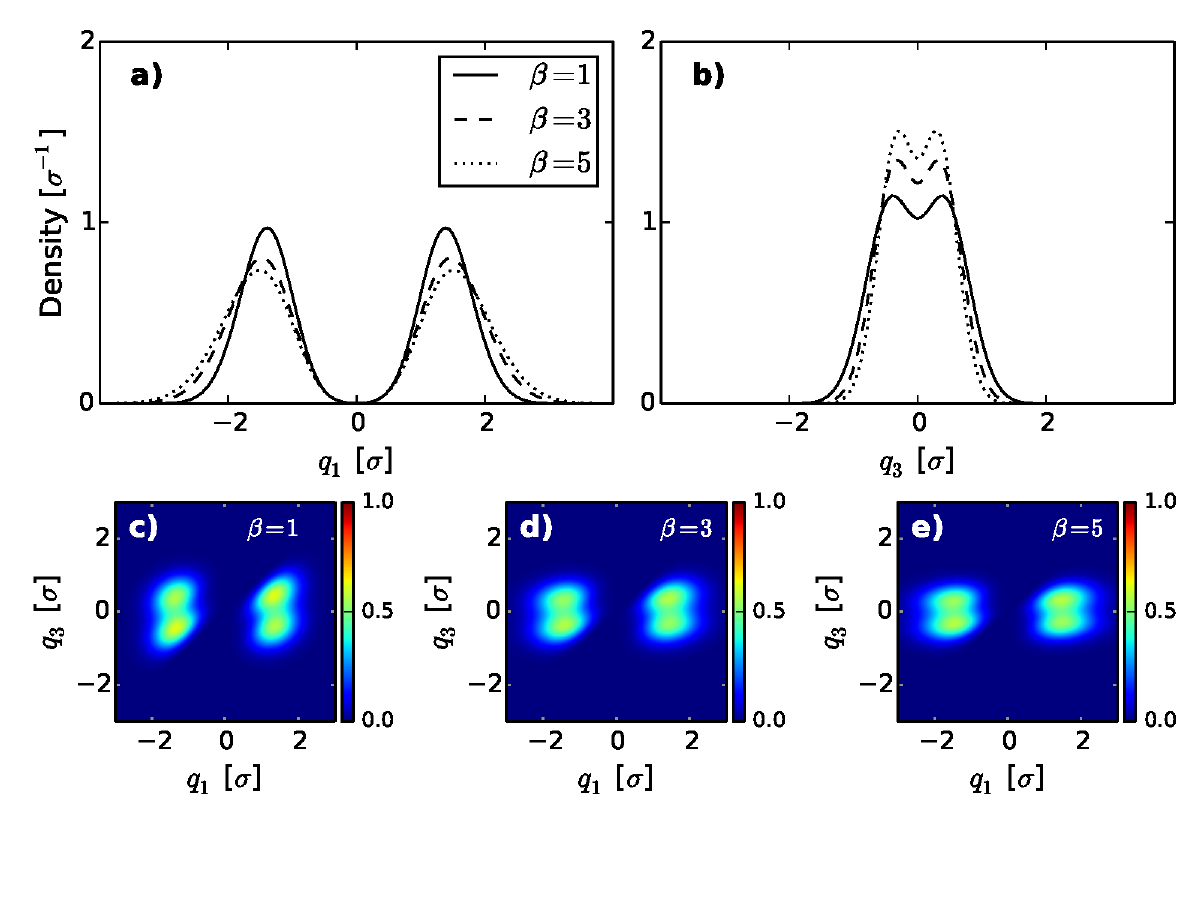}
\caption{(color online). The densities and pair-correlation functions for the ground state ($\uparrow\downarrow\downarrow\uparrow$ configuration) of the 2f+2f system. Panels {\bf a)} and {\bf b)} show the densities for the $\uparrow$ and $\downarrow$ particles, respectively, for $\beta=1+\epsilon,3$ and $5$. Panels {\bf c)}, {\bf d)} and {\bf e)} show the corresponding $\uparrow\downarrow$ pair-correlation functions.}
\label{density_2f+2f}
\end{figure}

\subsubsection{2f+2f} 
The 2+2 fermionic system with different masses is the final case we cover. 
The density for this system is shown in Fig.~\ref{density_2f+2f}. We see that the light component is pushed further to the side by the heavy fermions as the mass ratio increases. This is consistent with the findings of Ref.~\cite{pecak2015}. 
For $\beta\to\infty$, the heavy fermions occupy the ground and 1st excited single-particle states in the harmonic trap, whereas the light fermions sit at the double degenerate ground state of the harmonic oscillator with an impenetrable barrier in the middle. The energy of relative motion is then $9/2\hbar\omega$. Notice that for $\beta=1$ the ground state is six-fold degenerate since all configurations have the same energy, see, e.g., Ref.~\cite{volosniev2014a}. This degeneracy is, however, broken immediately if $\beta\neq1$, which is also found in the three-body case in Ref.~\cite{nielsjacob2015}. 
Thus, our results in Fig.~\ref{density_2f+2f} describe the ground 
state for $\beta=1+\epsilon$, where $\epsilon\to0$.

\subsection{Momentum distributions} 
\begin{figure}[t]
\centering
\includegraphics[width=\columnwidth]{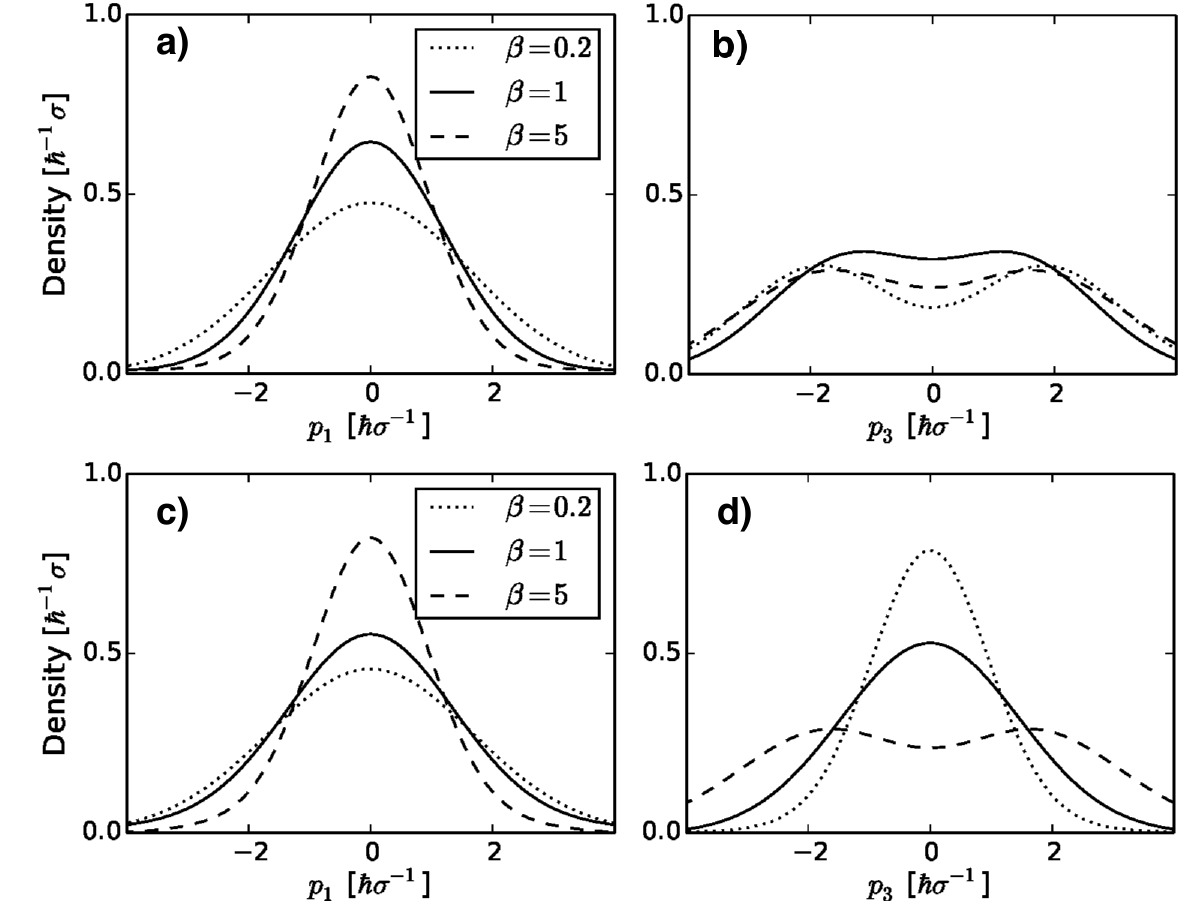}
\caption{(color online). Panel {\bf a)} shows the momentum distributions for the first species in the 2b+2b system as a function of the mass ratio $\beta$. 
Panel {\bf b)} is similar to {\bf a)} but for the 2f+2f system. Panels {\bf c)} and {\bf d)} show the distributions correspondingly for the bosons and fermions in the 2b+2f system.}
\label{momentum}
\end{figure}

Finally, we calculate the momentum distributions and present them in Fig.~\ref{momentum}. 
The momentum distributions for the 2b+2b system are shown in Fig.~\ref{momentum}{\bf a)}, that for 2f+2f in Fig.~\ref{momentum}{\bf b)}, and for 2b+2f in 
Fig.~\ref{momentum}{\bf c)} and {\bf d)}. The mass ratios are $\beta=0.2, 1$ and $5$. Fig.~\ref{momentum}{\bf a)} shows the distributions for only the first species in the 2b+2b system. Due to the symmetry in the two-component Bose system the corresponding distributions for the second species can be obtained from the same figure with $\beta\to1/\beta$ substitution. The distinctive feature is that as 
$\beta$ grows, the momentum distribution for the heavy particles spreads, while for the light particles it becomes more peaked. This is easily understood since the heavy particles become more localized in space.
Fig.~\ref{momentum}{\bf b)} shows the same as Fig.~\ref{momentum}{\bf a)} for the 2f+2f system. The momentum distribution for the fermions has its characteristic double peak. However, other than the peaks the same evolution is present as for the bosons.

In Fig.~\ref{momentum}{\bf c)} and {\bf d)} we show the momentum distributions for the ground state of the 2b+2f system. The bosons are described with Fig.~\ref{momentum}{\bf c)} while {\bf d)} shows the distribution for the fermions. Note that here the 2b+2f system is in the $\downarrow A A \downarrow$ configuration for $\beta=0.2$ and $\beta=1$, while for $\beta=5$ the ordering is $AA\downarrow\downarrow$. As seen in Fig.~\ref{momentum}{\bf d)} the momentum distribution reveals this transition as the fermionic two-peak structure appears for $\beta>\beta_c$.

\section{Conclusions}
We have studied short-range interacting particles in a one-dimensional harmonic 
trap in the limit 
where the interaction of a given pair is either of vanishing or of very large
(repulsive) strength. Using a rewriting of the $N$-body Schr{\"o}dinger equation, 
we have recast this problem in terms of hyperspherical coordinates for the case
of general masses of the particles. This is a particularly difficult case as 
it is generally non-integrable even in the homogeneous case. Recent advances in 
the theoretical description of strongly interacting equal mass particles in 
an external trap cannot be directly applied and this motivates the exploration 
of alternative approaches such as the one discussed here. As one may explore the
physics of mixtures of particles with different masses in present experiments it is 
crucial to have theoretical tools that apply to these cases as well.

To illustrate the formalism we investigate the case of four particles. This is 
a particularly interesting case as one may visualize the coordinate space of 
the problem in three-dimensional figures due to the decoupling of the total 
center-of-mass of the system. The prime case of interest was the 2+2 system 
with two particles of one kind and two particles of another. For a 
Bose-Bose mixture of two $A$ and two $B$, we find that when the $AB$ interaction
becomes strong the ground state configuration is $AABB\pm BBAA$ for
any mass ratio. In the case of a Fermi-Fermi 
mixture we find that the ground state is a $\uparrow\downarrow\downarrow\uparrow$
configuration for all \cite{footnote} $m_\downarrow> m_\uparrow$, i.e., that the heavier
fermions go to the center of the trap in the ground state. This is 
consistent with the study in Ref.~\cite{pecak2015} which includes a harmonic trap.
A similar separation of components in Fermi-Fermi mixtures has been seen in 
the homogeneous case with mass-imbalance in Ref.~\cite{cui2013} and 
in Ref.~\cite{fratini2014}.

For Fermi-Bose mixtures of two light $\uparrow$ fermions and 
two heavy $B$ bosons it is the $\uparrow BB\uparrow$ configuration that is the 
ground state. In the case with two light bosons, $A$, and two heavy fermions, $\downarrow$, we find an 
interesting transition in the ground state as a function of $\beta=m_\downarrow/m_A$ around 
a critical value of $\beta_c\simeq 1.3$.
For $1\leq \beta<\beta_c$ the configuration of the ground state is $\downarrow AA\downarrow$, 
while for $\beta> \beta_c$ this changes such that the ground state now has the 
configuration $AA\downarrow\downarrow$ (and its parity twin $\downarrow\downarrow AA$). 
Thus we have an example where the phase separation changes its character as function of 
the mass-imbalance. We are not aware of previous studies that have found this type 
of result. It would be interesting to extend to larger systems in order to see
how this critical value changes with particle number and/or system composition (number 
of fermions vs. number of bosons).

While our study here was based on transforming the problem into a form that can be 
handled numerically by standard methods through the use of hyperspherical coordinates,
the question of whether some exceptional analytically tractable cases might be found
of course remains. As noticed in the introduction, Ref.~\cite{olshanii2015} has presented
an exact solutions for the case of four particles in a hard-wall box in one dimension when 
the masses have specific integer ratios. This was done by a symmetry analysis of the 
coordinate space of the system and the use of the Bethe ansatz. The formalism we 
present here is different and the harmonic oscillator confining the system of course
changes the boundary conditions at large distances. Nevertheless, the large degree of 
symmetry that the harmonic potential brings, and the elegant way in which this implies 
separation in hyperspherical coordinates between the radial and angular parts provides 
some optimism that one may be able to find some exceptional analytical solutions also 
in the present case for certain mass ratios. We leave this question for future work. 

\appendix

\section{3+1 system}\label{app:A}
\begin{figure}
\centering
\includegraphics[width=\columnwidth]{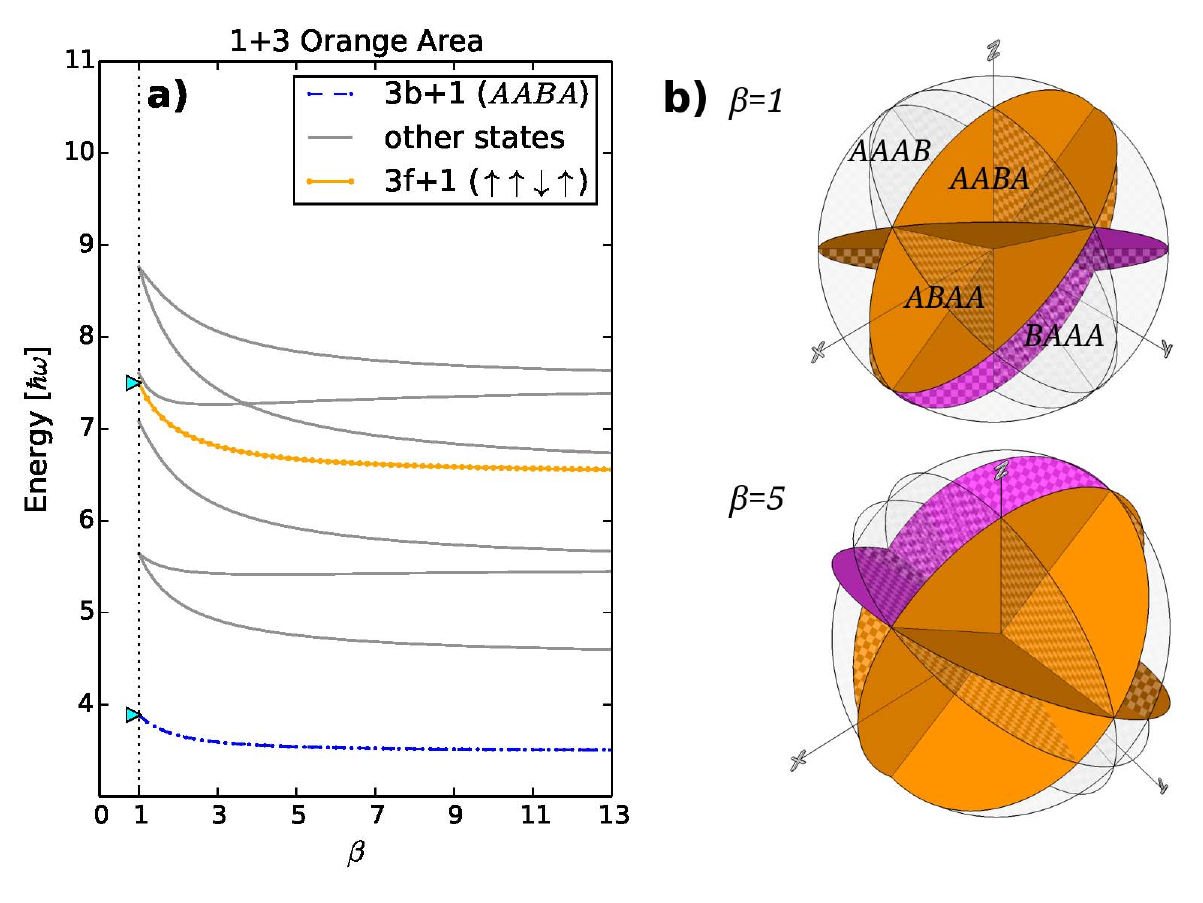}
\caption{ (color online). {\bf a)} The energy spectrum for the 3+1 system for the configurations depicted in the inset. The (yellow) dotted curve corresponds to the 3f+1 system. The (blue) dot-dashed line is for the 3b+1 system. {\bf b)} The geometry of the relative motion space for the 3+1 system for $\beta=1$ and $\beta=5$.}
\label{space_3+1}
\end{figure}

The steps discussed for the 2+2 system in the main text can be taken also in the 3+1 case. We discuss this briefly for the 3f+1 and 3b+1 systems. Following the same route as for the 2+2 systems we end up with the geometry of the relative motion space depicted in 
Fig.~\ref{space_3+1}{\bf b)}. Using a further coordinate transformation for the upper (orange) $AABA$ region we end up with simple boundary conditions which suggest the use of the Fourier transform from above. This allows us to obtain the energy plot in 
Fig.~\ref{space_3+1}{\bf a)}. However, as seen in Fig.~\ref{space_3+1}{\bf b)} for $\beta=5$, the upper (orange) region, even though, is symmetric, is not necessary the largest. This implies that there is a possibility that the ground state is to be found in one of the other regions in the figure (most likely the (orange) region nearest to the $xy$ plane). Even though these other regions can be easily addressed, they do not allow a simple use of the Fourier transform used for the 2+2 system, so we left them for future studies.

When it comes to the 3b+1 system, we no longer have the irregular division of the (orange) area, because the checkerboard patterned planes are no longer relevant, and therefore one can easily solve the problem and find a state in the (orange) area. However, this area is clearly not the largest area, and therefore it most likely corresponds to the excited states of the 3b+1 system. The largest area in this case is the (purple) area along the negative $y$-axis. Again this region requires a treatment different from the one we used for the 2+2 system. Therefore, we set this aside for future investigations.

\section{Angular equation}\label{app:B}
In order to find $f$ we solve the angular equation together with the boundary conditions given by the arguments of the delta-function potentials. For this we perform two transformations: i) $a=\cos\phi~\tan\theta$, $
b=\sin\phi~\tan\theta$, and ii) $\lambda=a~\sin\xi - b~\cos\xi$, $\gamma=a~\sin\xi + b~\cos\xi$. It is worthwhile noting that in terms of the original coordinates, $q_i$, these variables are given as
\begin{align*}
\lambda=\sqrt{1+\frac{\beta}{\alpha}}\cdot\frac{ \sqrt{{\alpha}/{\beta}}({q_1-q_2})\sin\xi-({q_3-q_4})\cos\xi}{q_1+q_2-q_3-q_4} ,\\
\gamma=\sqrt{1+\frac{\beta}{\alpha}}\cdot\frac{ \sqrt{{\alpha}/{\beta}}({q_1-q_2})\sin\xi+({q_3-q_4})\cos\xi}{q_1+q_2-q_3-q_4}.
\end{align*}
These transformations lead to very simple boundary conditions, namely the wave function vanishes if $\lambda=\pm1$ or $\gamma=\pm1$. The corresponding wave function
should be found from the angular equation, which now reads
\begin{widetext}
\begin{align}
(1+\lambda^2)\frac{\partial^2 f}{\partial \lambda^2}+(1+\gamma^2)\frac{\partial^2 f}{\partial \gamma^2}+(2\lambda\gamma-2\cos(2\xi))\frac{\partial^2 f}{\partial \lambda\partial \gamma}+2\lambda\frac{\partial f}{\partial \lambda}
+2\gamma\frac{\partial f}{\partial \gamma}= 
\frac{\tau(\tau+1)\sin(2\xi)^2 f}{\sin(2\xi)^2+\lambda^2+\gamma^2+2\lambda\gamma \cos(2\xi)}.
\label{eq:angf}
\end{align}
\end{widetext}

To solve this equation we expand the wave function in the Fourier series which obeys the boundary conditions
\begin{equation} 
f(\lambda,\gamma)= \sum_{n,m} C_{n,m}~\sin\left[\frac{\pi n}{2}(\lambda-1)\right]
\sin\left[\frac{\pi m}{2}(\gamma-1)\right]\;.
\end{equation} 
Note that the function is defined on a square $-1<\lambda<1, \; -1<\gamma<1$. By allowing $n$ and $m$ to run up to some maximum value, say $n_{max}=m_{max}=20$, one ends up with a matrix equation for the coefficients $C_{n,m}$. Indeed, if we first insert the Fourier decomposition into Eq.~(\ref{eq:angf}) and then use the fact that the basis is orthonormal, i.e., $\int_{-1}^{1} \sin[{\pi n}/{2}(\lambda-1)]~\sin[{\pi m}/{2}(\gamma-1)]=\delta_{nm}$ we end up with a simple matrix eigenvalue problem whose eigenvectors and eigenvalues are $C_{n,m}$ and $\tau(\tau+1)$ respectively. After the set $C_{n,m}$ is established we have all information about the system.\\

Finally we mention that to calculate the density profiles we split the normalization constant such that
\begin{align*}
\int\left|R(\rho)\right|^2 \frac{\mu}{\sqrt{M}}\mathrm{d}\rho=1,\\
\int|f|^2 \sin(\theta)\mathrm{d}\phi \mathrm{d}\theta=1.
\end{align*}
To find the latter equation in terms of $a,b$ and then $\lambda,\gamma$ one should use that
\begin{align*}
&\sin(\theta)\mathrm{d}\phi \mathrm{d}\theta\rightarrow \frac{1}{(1+a^2+b^2)^{3/2}}\mathrm{d}a \mathrm{d}b
\rightarrow \\&{\sin(2\xi)^2}/{[\sin(2\xi)^2+\lambda^2+\gamma^2+2\lambda\gamma \cos(2\xi)]^{3/2}} \mathrm{d}\gamma \mathrm{d}\lambda.
\end{align*}

\section{The density and pair-correlation function}\label{app:C} 
In our analytical results we use the four-body wave function $\psi(q_1,q_2,q_3,q_4)$ to obtain the density of the system,
\begin{align}
n_A(q_1)=\int{|\psi|^2\mathrm{d}q_{2}\mathrm{d}q_{3}\mathrm{d}q_{4}},\\
n_B(q_3)=\int{|\psi|^2\mathrm{d}q_{1}\mathrm{d}q_{2}\mathrm{d}q_{4}}.
\end{align}
For the 2b+2b system we define the pair-correlation function for an $AB$ pair as
\begin{equation}
n_{AB}(q_1,q_3)=\int{|\psi|^2 dq_{2} dq_{4}},
\end{equation}
similarly we define the pair-correlation functions for the other systems discussed in the paper.
To find the momentum distribution we first define the Fourier transform
of the wave function:
\begin{align*}
\psi(p_1&,\dots,p_{4})=\left(\frac{1}{\sqrt{2\pi}}\right)^{4}\\
&\int_{all~space}{\psi(q_1,\dots,q_{4})e^{ip_1q_1}\dots e^{ip_{4}q_{4}}}dq_1\dots dq_{4}
\end{align*}
where $p_i$ is the momentum of the particle. 
This allows us to obtain the momentum distributions for the two species,
\begin{align*}
n_1(p_1)=\frac{1}{2\pi}\int_{all~space}{\psi^* (q_1,q_2,q_3,q_4)\psi(\tilde{q}_1,q_2,q_3,q_4) \cdot }\\ 
{e^{ip_1(q_1-\tilde{q}_1)}~dq_1d\tilde{q}_1dq_2dq_3dq_4}\\
n_2(p_3)=\frac{1}{2\pi}\int_{all~space}{\psi^* (q_1,q_2,q_3,q_4)\psi(q_1,q_2,\tilde{q}_3,q_4) \cdot }\\ 
{e^{ip_1(q_3-\tilde{q}_3)}~dq_1dq_2dq_3d\tilde{q}_3dq_4}
\end{align*}

\begin{acknowledgments}
We thank Xiaoling Cui and Tomasz Sowi{\'n}ski for feedback on the manuscript. 
This work was funded by the 
Danish Council for Independent Research DFF Natural Sciences and the 
DFF Sapere Aude program. 
A.G.V. acknowledges partial support by Helmholtz Association under contract HA216/EMMI.
\end{acknowledgments}


\begin{thebibliography}{99}
\bibitem{footnote} For the convenience of the reader in this section we shall use self-explanatory subscripts for masses.
\end{thebibliography}
\end{document}